\documentclass[12pt]{article}
\usepackage{amsmath,amsfonts, amssymb,graphicx,geometry,authblk,color,soul,comment,hyperref} 
\usepackage{subfigure}

\usepackage{natbib}
\setcitestyle{round}

\newcommand{\comm}[1]{}

\title{Iterated learning and multiscale modeling of history-dependent architectured metamaterials}
\author{Yupeng Zhang and Kaushik Bhattacharya}
\affil{California Institute of Technology, Pasadena CA 91125}
\date{\today}

\begin{document}

\maketitle
	
\begin{abstract}

Neural network based models have emerged as a powerful tool in multiscale modeling of materials.  One promising approach is to use a neural network based model, trained using data generated from repeated solution of an expensive small scale model,  as a surrogate for the small scale model  in application scale simulations.  Such approaches have been shown to have the potential accuracy of concurrent multiscale methods like FE$^2$, but at the cost comparable to empirical methods like classical constitutive models or parameter passing.  A key question is to understand how much and what kind of data is necessary to obtain an accurate surrogate.  This paper examines this question for history dependent elastic-plastic behavior of an architected metamaterial modeled as a truss.  We introduce an iterative approach where  we use the rich arbitrary class of trajectories  to train an initial model, but then iteratively update the class of trajectories with those that arise in large scale simulation and use transfer learning to update the model.  We show that such an approach converges to a highly accurate surrogate, and one that is transferable.
\vspace{\baselineskip}

\end{abstract}

\section{Introduction}

The behavior of materials is determined by diverse phenomena that occur at a range of diverse time and length scales.  Multiscale modeling exploits the separation of scales to map the complexity into a hierarchy of distinct phenomena at distinct scales, and then integrates them back through a series of pairwise interaction.    This framework has its roots in homogenization theory and coarse-graining, and has provided important insights in a variety of phenomena.
See for example \cite{van_der_giessen_roadmap_2020} for a review, \cite{fuhg_model-data-driven_2021}, \cite{xue_simulation_2022}, \cite{vasoya_modeling_2023} for discussions.

A key challenge in this framework is the passage of information from one scale to another.  There are accurate concurrent multiscale approaches like FE$^2$, but these are prohibitively computationally expensive.  At the other end, there are empirical approaches like parameter passing, but these have limited fidelity.  In recent years, machine learning, where the small scale model is replaced with a neural network-based surrogate that is trained by data generated off-line by repeated solution of the small scale model, has emerged as a promising solution 
\citep[e.g.,][]{bessa_framework_2017,vlassis_sobolev_2021,gupta_accelerated_2023,baek_neural_2024}.
Specifically, recurrent neural networks and its variations have emerged as a common approach to describe history dependent behavior \citep[e.g.,][]{lefik_artificial_2009,ghavamian_accelerating_2019,mozaffar_deep_2019,wu_recurrent_2020,logarzo_smart_2021,wu_recurrent_2022}.

Inspired by the successful use of internal variable theories in continuum mechanics, \cite{liu_learning_2023} proposed a variation of this idea, recurrent neural operators (RNO).   While they have a structure inspired by internal variable theories, the internal or state variables are not identified {\it a priori}, but learnt from the data as a part of the training.  These are neural operators that are formulated in a time continuous manner, and thus can be used independent of the time discretization of either the training data or the application.  They applied this to elasto-viscoplasticity, and showed that an RNO can be trained to accurately learn the overall response of a polycrystalline unit cell with a remarkably small number of internal variables.  They also showed that the RNO can be used in a finite element calculation, thereby achieving FE$^2$ accuracy at the cost of classical simulations.  Other applications of this idea include porous media flow \citep{karimi_learning-based_2023} and viscoelasticity \citep{bhattacharya_learning_2023}.

The promise of this approach raises the question of understanding how much and what kind of data is necessary to obtain an accurate surrogate.  There are mathematical analysis of neural operators that provide universal approximation theorems \citep{kovachki_neural_2023}.  There is also a characterization of the data necessary for systems governed by elliptic partial differential equations (static problems) \citep{boulle_learning_2023}.  However, this issue has not be examined in the setting of history dependent problems, and is the primary motivation for this work in the setting of visco-plasticity.  

One may take a statistical point of view and consider an Ornstein-Uhlenbeck process \citep{uhlenbeck_theory_1930}.  However, these systems are dissipative and therefore filter out high frequency oscillations.  Further applications typically lead to relatively few changes of loading direction.  This motivated previous work to consider data generated from trajectories where each component of deformation gradient is smooth but arbitrarily change rate (positive and negative) at a finite number of instances \citep{liu_learning_2023,bhattacharya_learning_2023,karimi_learning-based_2023}.  This corresponds to a large variety of non-proportional loads with a range of strains and strain-rates.  {\it A posteriori} analysis revealed an error on trajectories in applications \citep{karimi_learning-based_2023} that is reasonable, but significantly higher than those in the original class of trajectories.  This suggests that the trajectories used for training may not be optimal in application.

In this work, we examine this question recalling that the primary motivation for the creation of a machine learnt surrogate is its use in larger scale models.  So, we view the question of data generation and accuracy from the point of its efficacy in large scale simulations.  We propose an approach where we use the rich arbitrary class of trajectories as in previous work to train an initial model, but then iteratively update the class of trajectories with those that arise in large scale (finite element) analysis and use transfer learning 
\citep[e.g.,][for reviews]{pan_survey_2010,zhuang_comprehensive_2021}
to update the model.  We show that such an approach converges to a highly accurate RNO surrogate, and one that is transferable.  So this work provides insight that may be the basis of a systematic examination of this question.

A second motivation of the work is to examine the efficacy of RNOs in multiscale modeling of architected metamaterials.  Architected metamaterials show a promise in a variety of applications due to the ability to engineer them with a range of unusual mechanical properties and to synthesize them using additive manufacturing
\citep[e.g.,][]{meza_resilient_2015,vyatskikh_additive_2018,kochmann_multiscale_2019,xia_electrochemically_2019,jin_mechanical_2023}.  Multiscale modeling with machine learning has been used to bridge the length-scales that may arise, but much of the work has concerned either linear behavior characterized by unusual dispersion or nonlinear elastic behavior \citep[e.g.,][]{le_computational_2015,white_multiscale_2019}.  We study history dependent visco-plasticity.  Conversely, prior learning on plasticity consider continuum composite media and polycrystals.   Architected metamaterials have unusual plastic behavior since compaction can lead to non-volume preserving plastic strains.   We show that RNOs provide an effective representation of their effective behavior with a relatively small number of internal variables.  

We consider a large periodic truss in two dimensions as our representative architected material.  We model it in a multiscale setting where we consider a continuum at the large scale (large compared to the unit cell), and whose constitutive behavior is computed from the overall response of the unit truss.  We consider a random truss in finite deformation with a large number of elastic plastic links as our unit cell.  We introduce the truss and the model in Section \ref{sec:truss}, and describe the computation of its effective behavior.  We introduce a recurrent neural operator (RNO) as a surrogate in Section \ref{sec:rno}, and describe its training and performance.  
We turn to macroscopic simulations using the trained RNO as a material model in Section \ref{sec:FEM}.  The core contribution of the work, iterated learning and its evaluation is presented in Section \ref{sec:transferLearning}.  We conclude in Section \ref{sec:conc} with a discussion and perspective

\section{Microscale Model and Effective Behavior} 
\label{sec:truss}

\begin{figure}
	\begin{center} 
	{\includegraphics[width=6in]{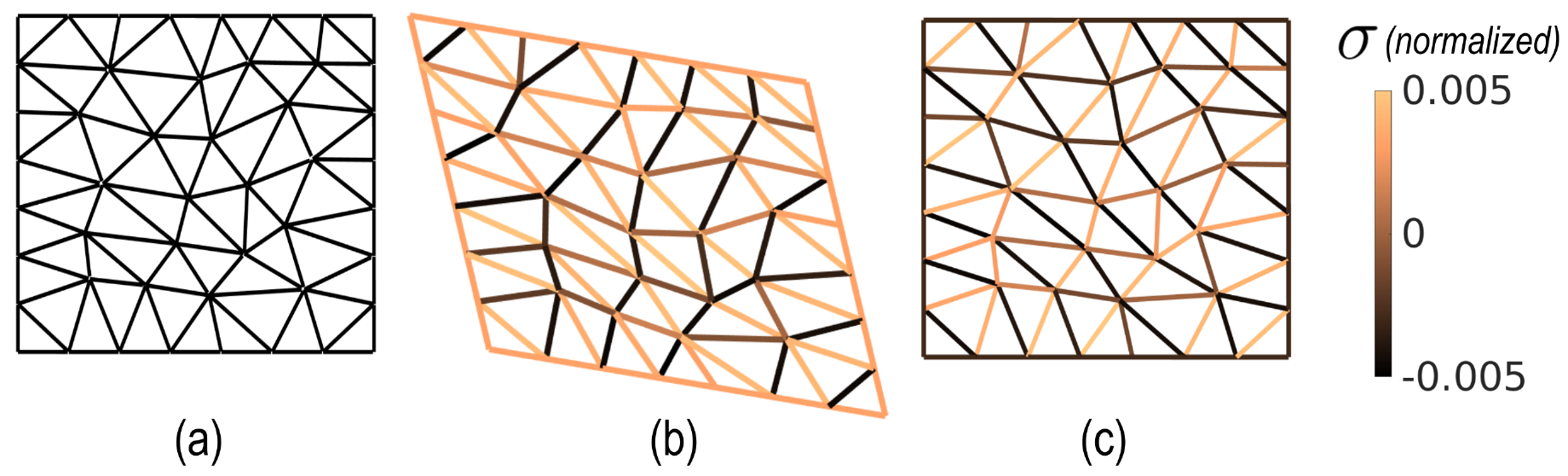}}
	\caption{(a) The unit truss consisting of 113 links and 48 nodes occupying a unit square.
		(b) The current configuration of truss subjected to a homogenous boundary condition simple shear S-- at deformation gradient [1,-0.2;-0.2,1] in Section \ref{sec:simple_modes}; 
		(c) The current configuration of truss after loading-unloading of simple shear S-- in Section \ref{sec:simple_modes}.
	    The contours in (b) and (c) denote the scalar stress in
        each one dimensional link. \label{fig:truss}}
    \end{center}
\end{figure}

\subsection{Elastic viscoplastic truss}
\label{sec:constitutive}

Consider a truss consisting of $L$ linear links with initial stress-free lengths $\{L^0_l\}$ and $N$ (pin-jointed) nodes -- see Fig.~\ref{fig:truss}(a) for example.  The connectivity of the truss is described by the connectivity matrix, 
\begin{equation}
B_{ln} = \begin{cases} 1& n \text{ largest node number connected to link } l, \\
-1& n \text{ smallest node number connected to link } l, \\ 
0 & \text{else}. \end{cases}
\end{equation}
Let $y_n$ denote the current position of the $n^\text{th}$ node.  The current length of the $l^\text{th}$ element is $L_l = \left| \sum_n B_{ln}y_n \right|$.  The change of length in the $l^\text{th}$ link is $U_l = L_l - L_l^0$.   We assume that the links are elastic-plastic, and denote the plastic change of length of the $l^{th}$ link to be $U_l^\text{pl}$.  The total, plastic and elastic strains in the $l^\text{th}$ link are
\begin{equation}
\epsilon_l = \frac{U_l}{L_l^0}, \quad \epsilon_l^\text{pl} = \frac{U_l^\text{pl}}{L_l^0}, \quad \epsilon_l^\text{el}= \frac{U_l^\text{el}}{L_l^0} = \epsilon_l - \epsilon_l^\text{pl}
\end{equation}
respectively.  We also define the accumulated plastic strain in the $l^\text{th}$ link as
\begin{equation} \label{eq:plastic work}
\alpha_l(t) = \int_0^t \dot \alpha_l (\tau) d \tau, \quad  \dot \alpha_l = |\dot \epsilon_l^\text{pl}|.
\end{equation}

We fix the nodes in the set ${\mathcal N}_y$ at position $\bar y_n, n \in {\mathcal N}_y$, and apply (possibly zero) forces $\bar f_n, n \in {\mathcal N}_F$ on the remaining nodes ${\mathcal N}_f$.  The total stored energy in the system is given by 
\begin{equation}
{\mathcal E} (\{y_n\},\{\epsilon^\text{pl}_l \}, \{\alpha_l \} ) = \sum_l \left( W^\text{el}( \epsilon^\text{el}_l) + W^\text{pl} (\alpha_l) \right) L_l^0 - \sum_{n \in {\mathcal N}_f} \bar f_n y_n
\end{equation}
and the rate of dissipation by 
\begin{equation}
{\mathcal D} ( \{ \dot \alpha_l \} ) = \sum_l D(\dot \alpha_l) L_l^0 
\end{equation}
where
\begin{align}
W^\text{el}( \epsilon) &= \frac{K}{2} \epsilon^2, \\
W^\text{pl}(\alpha) & = f_0 \left( \alpha+\frac{\epsilon^p_0}{N+1}\left( \frac{\alpha}{\epsilon^p_0} \right)^{N+1} \right), \\
D(\alpha) &=  f_0  \left(\frac{\dot{\epsilon}^p_0}{m+1}\right) \left( \frac{\dot{\alpha} }{\dot{\epsilon}^p_0} \right)^{m+1} 
\end{align}
are stored elastic energy density,  stored plastic work density and the dissipation rate density, respectively.  $f_0$ is the initial yield force $f_0=K \epsilon_0$, $K$ the truss spring constant,
$N$ the strain hardening exponent, $\epsilon_0^p$ the reference plastic strain, $m$ the rate hardening exponent and $\dot \epsilon_0^p$ the reference plastic strain rate.

For a (possibly time-dependent) set of prescribed displacements $\bar y_n, n \in {\mathcal N}_y$, and prescribed forces $f_n, n \in {\mathcal N}_F$, we find the evolution of the kinematic variables $V = (\{y_n\},\{\epsilon^\text{pl}_l \}, \{\alpha_l \} ) $ as
\begin{equation} \label{eq:evol}
0 \in \partial_V {\mathcal E} + \partial_{\dot V} {\mathcal D}
\end{equation}
subject to (\ref{eq:plastic work}). Since $\dot \alpha \ge 0$ by definition, the dissipation rate ${\mathcal D}$ is not smooth and this results in a differential inclusion (corresponding to a Kuhn-Tucker condition) instead of an equality.  See \ref{appendix_numerical} for details.

For future use, we note that the reaction force at a node $n \in {\mathcal N}_y$ is given by 
\begin{equation}
f_n = \frac{\partial \mathcal E}{\partial y_n} = \sum_l  K \epsilon_l^\text{el} \hat v_{ln}
\end{equation}
where $\hat v_{ln} = B_{ln} \sum_m B_{lm} y_m / |  \sum_m B_{lm} y_m |$ is the unit vector along the link $l$ pointing towards node $n$.  The second equality follows by a repeated use of the chain rule.   

\subsection{Effective behavior of a square truss}
\label{sec_CauchyStress}

We now consider a truss occupying a square $\Omega$ as shown in Fig.~\ref{fig:truss}(a).   We take ${\mathcal N}_y$ to be all the nodes on the boundary $\partial \Omega$.  We prescribe a homogenous boundary condition 
\begin{equation} \label{eq:bc}
\bar{y}_n(t) = F(t) x_n, \quad n \in {\mathcal N}_y
\end{equation}
corresponding to an average deformation gradient (history) $F(t)$. A typical example is shown in Fig.~\ref{fig:truss}(b).  We assume zero applied forces $f_n=0, n \in {\mathcal N}_f$.  

We now seek to define an average stress for the entire truss.  To do so, we define the energy density per unit area 
\begin{equation}
W(F) = \min_{y_n \in {\mathcal N}_f} \frac{1}{A_\Omega} \ {\mathcal E}(\{y_n\},\{\epsilon^\text{pl}_l \}, \{\alpha_l \} )
\end{equation}
subject to the homogeneous boundary conditions (\ref{eq:bc}). ${A_\Omega}$ is the area in the reference configuration.
We can now define Piola-Kirchhoff stress as
\begin{equation}
P_{ij} = \frac{\partial W}{\partial F_{ij}} =  \frac{1}{A_\Omega}  \sum_{n \in {\mathcal N}_y} 
\frac{\partial \mathcal E}{\partial (y_n)_k} \frac{\partial (y_n)_k}{\partial F_{ij}} =  \frac{1}{A_\Omega}  \sum_{n \in {\mathcal N}_y}  (f_n)_i (x_n)_j.
\end{equation}
Above, we only have the sum over the boundary nodes since the other nodes are in equilibrium by definition of $W$.  It follows that the Cauchy stress $\sigma = (\det F)^{-1} P F^T$ is given by 
\begin{equation} \label{eq:cauchy}
\sigma_{ij}  = \frac{1}{A_\Omega  \det F} \sum_{n \in {\mathcal N}_y}  (f_n)_i (y_n)_j =  \frac{1}{A_\Omega \det F}\sum_{n \in {\mathcal N}_y}  (f_n)_j (y_n)_i.
\end{equation}
Note that we divide by area and these stresses are defined as force per length since we are in two dimensions\footnote{In three dimensions, we would divide by volume and the stress would have its usual units.}.  Alternately, we may regard this as stress per unit thickness in plane strain.  The final statement above regarding the symmetry of the Cauchy stress follows from the frame-indifference of $\mathcal E$.

We are now in a position to define the overall or effective behavior. Given a deformation gradient history or deformation gradient trajectory $F: [0,t] \to {\mathbb F}^{2 \times 2}$, we solve the governing equations described in Section \ref{sec:constitutive} for the position of the nodes and plastic deformation of the nodes subject to the homogeneous boundary conditions (\ref{eq:bc}).  We compute the Cauchy stress $\sigma(t)$ from (\ref{eq:cauchy}).  The effective or overall behavior of the truss is the map from the deformation gradient history to the current Cauchy stress:
\begin{equation} \label{eq:effective}
{\mathcal F}: \{F(\tau): \tau \in [0,t]\} \to \sigma(t).
\end{equation}

\subsection{Typical results}
\label{sec_SampleResults}

We generate a random architected truss material as follows.  We consider a square domain with side-length $a$ and generate an equally spaced set of nodes ${\mathcal N}_y$ on the boundary.  We then generate an interior set of nodes ${\mathcal N}_f$ by sampling random points in the interior of $\Omega$ (and discarding points that are closer than
$a\sqrt{\lambda/|{\mathcal N}_f|}$ some $0 < \lambda < 1$ and $|{\mathcal N}_f|$ denotes the size of the set.  We then generate the links by Delaunay triangulation of the nodes ${\mathcal N} = {\mathcal N}_f \cup {\mathcal N}_y$.  A typical example with 48 nodes and 113 links is shown in Fig.~\ref{fig:truss}.    We present all results in this paper for this particular example unless otherwise specified.

We non-dimensionalize all lengths by the side-length $a$ of the domain, all times by the total duration of loading $t_0$ and all forces by the truss spring constant $K$.   We use the parameters shown in Table \ref{tab:param} in the rest of the paper unless otherwise specified.

We solve for the equations using MATLAB.  We have verified the numerical method against analytical solutions for a four link truss.

\begin{table}
\centering
\caption{Table of non-dimensional parameters \label{tab:param}}
\begin{tabular}{cccccc}
\hline
 $\epsilon_0$ & $\epsilon^p_0$ & $N$ & $\dot{\epsilon}_0 t_0$ & $m$ \\
 0.002 & 0.1 & 0.3 & 10$^{5}$ &0.4\\
\hline
\end{tabular}
\end{table}

\subsubsection{Simple deformation modes}
\label{sec:simple_modes}

We consider five simple deformation modes:
\begin{itemize}
\item Horizontal tension (HT): uniaxial extension along $x_1$:
$$F: \begin{pmatrix} 1 & 0 \\ 0 & 1 \end{pmatrix} \to \begin{pmatrix} 1.2 & 0 \\ 0 & 1 \end{pmatrix} \to \begin{pmatrix} 1 & 0 \\ 0 & 1 \end{pmatrix}$$
\item Horizontal compression (HC): uniaxial compression along $x_1$:
$$F: \begin{pmatrix} 1 & 0 \\ 0 & 1 \end{pmatrix} \to \begin{pmatrix} 0.8 & 0 \\ 0 & 1 \end{pmatrix} \to \begin{pmatrix} 1 & 0 \\ 0 & 1 \end{pmatrix}$$
\item Vertical tension (VT): uniaxial extension along $x_2$:
$$F: \begin{pmatrix} 1 & 0 \\ 0 & 1 \end{pmatrix} \to \begin{pmatrix} 1 & 0 \\ 0 & 1.2 \end{pmatrix} \to \begin{pmatrix} 1 & 0 \\ 0 & 1 \end{pmatrix}$$
\item Vertical compression (VC): uniaxial compression along $x_2$:
$$F: \begin{pmatrix} 1 & 0 \\ 0 & 1 \end{pmatrix} \to \begin{pmatrix} 1 & 0 \\ 0 & 0.8 \end{pmatrix} \to \begin{pmatrix} 1 & 0 \\ 0 & 1 \end{pmatrix}$$
\item Shear + (S+): shear in the $+$ sense:
$$F: \begin{pmatrix} 1 & 0 \\ 0 & 1 \end{pmatrix} \to \begin{pmatrix} 1 & 0.2 \\ 0.2 & 1 \end{pmatrix} \to
\begin{pmatrix} 1 & 0 \\ 0 & 1 \end{pmatrix}$$
\item Shear -- (S--): shear in the $-$ sense:
$$F: \begin{pmatrix} 1 & 0 \\ 0 & 1 \end{pmatrix} \to \begin{pmatrix} 1 & -0.2 \\ -0.2 & 1 \end{pmatrix} \to
\begin{pmatrix} 1 & 0 \\ 0 & 1 \end{pmatrix}$$
\end{itemize}
Here, the notation $A \to B$ means that the deformation gradient changes from $A$ to $B$ linearly.
Each arrow occurs over the same time duration.

\begin{figure}[t]
	\begin{center}
	{\includegraphics[width=5.5in]{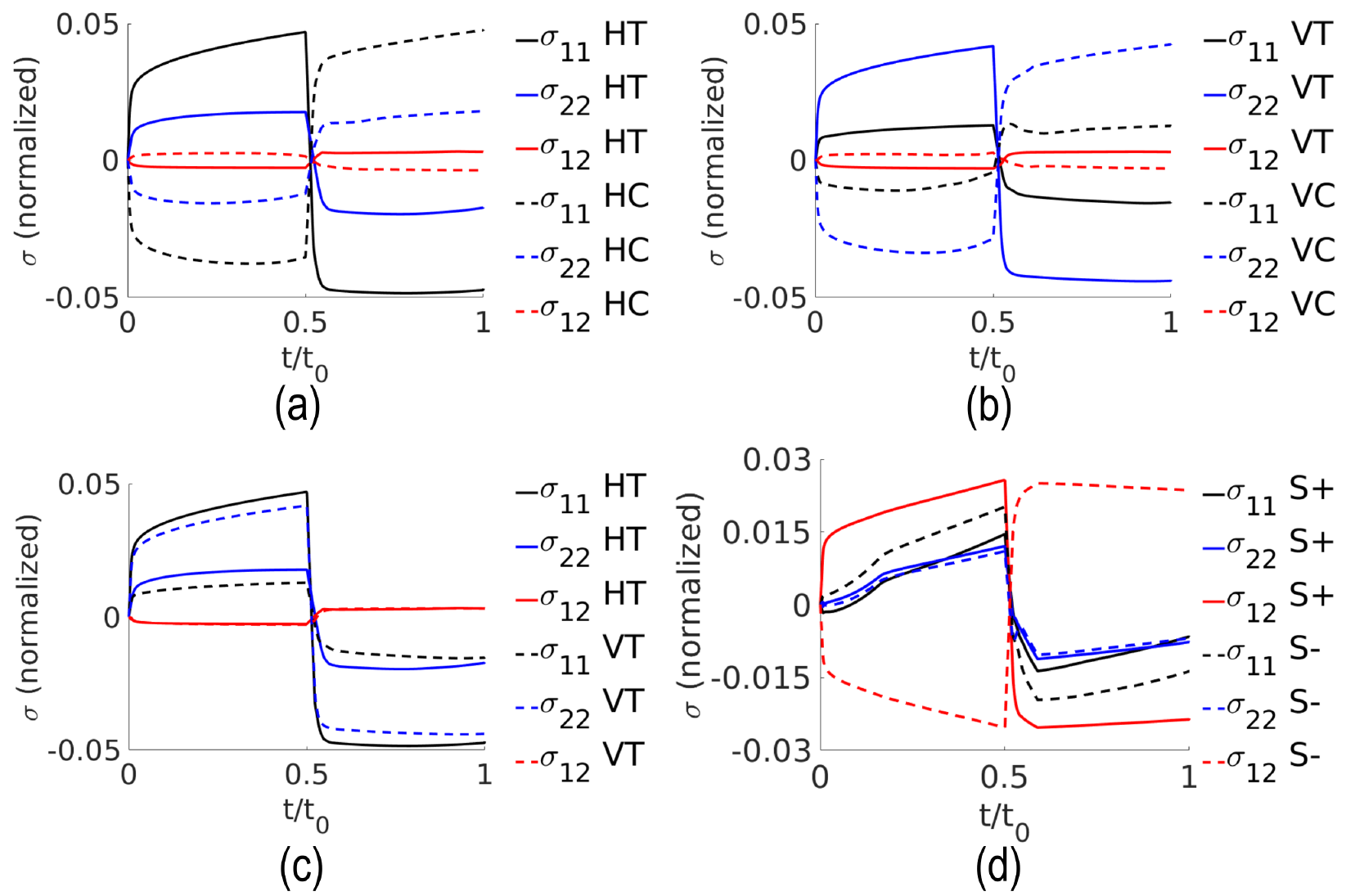}}
	\caption{Effective Cauchy stress vs. time for the simple deformation modes in Section \ref{sec:simple_modes}.}
	\label{fig:hom}
	\end{center}
\end{figure}

The results are shown in Fig.~\ref{fig:hom}.  We start wtih HT where the effective stress evolution is shown in Fig.~\ref{fig:hom}(a).  Following the normal stress $\sigma_{11}$, we see that the overall behavior is initially elastic and then gradually becomes plastic with hardening.  The yield is gradual as a result of averaging over many links.   When the applied extension is reversed, the structure initially unloads elastically but then builds up residual compressive stress due to uneven yield in the links.  It eventually yields in compression, but the behavior is different.  This is an emergence of the Bauschinger effect \citep{lubliner_plasticity_1990}.  Recall from Section \ref{sec:constitutive} that our links do not have kinematic hardening and so do not display the Bauschinger effect.  Still, we see the Bauschinger effect in the overall behavior due to an uneven distribution of plastic strains among the various links (Fig.~\ref{fig:truss}(b)).  We also see that the shear stress is non-zero, though small: this is a manifestation of the small anisotropy in the medium (see more detailed discussion below).

We now compare HT and HC, also  in Fig.~\ref{fig:hom}(a).  We find that the responses are different.  This is an emergence of tension-compression asymmetry \citep[e.g.,][]{ezz_tensioncompression_1982}.  Recall from Section \ref{sec:constitutive} that our links are symmetric.  Still we see asymmetry due to the effects of geometry in our finite deformation formulation.  Following the normal stress $\sigma_{11}$, we see a slight softening as deformation proceeds: this is again a result of geometry.

Fig.~\ref{fig:hom}(c) compares HT with VT.  We see similar, but slightly different behavior.  Since the truss is generated from randomly chosen points, there is a slight anisotropy, and this results in differences between HT and VT.  

Finally, we study shear in Fig.~\ref{fig:hom}(d).  We see elastic, plastic and hardening behavior in both applied shears.  An interesting observation is the emergence of tensile normal stresses independent of the direction of shear.  This is the Kelvin-Poynting effect, which in hyperelastic materials, is an even function of the shear strain \citep{poynting_pressure_1909}.  Finally, note that there is significant internal stress even when the applied strain is returned to zero, Fig.~\ref{fig:truss}(c).

\begin{figure}
	\begin{center}
		{\includegraphics[width=6.5in]{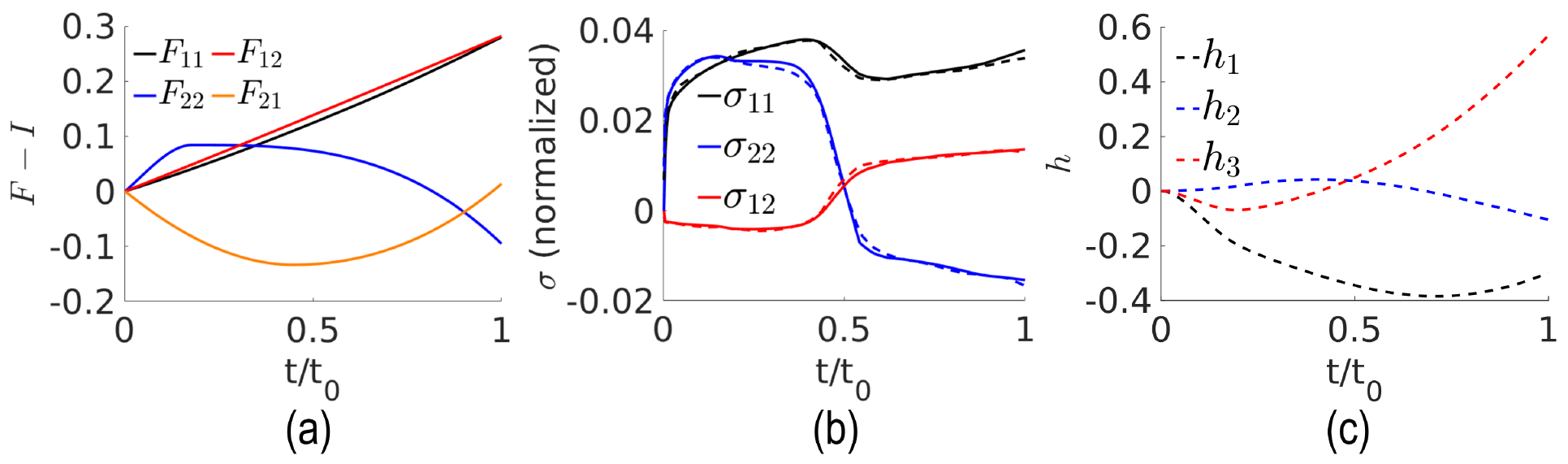}}
		\caption{Example of random deformation gradient trajectories (a),
			and the corresponding homogenized Cauchy stress (b) and internal variables (c).
			The solid lines are the data and the dashed lines denote the responses obtained from $RNO_1$
			trained in Section \ref{sec:rno}.}
		\label{fig:randomFS}
	\end{center}
\end{figure}

\subsubsection{Random deformation trajectories}
\label{sec_HomoF}

In the next section, we will seek to find a neural network approximation for the overall or effective behavior (\ref{eq:effective}) of the truss.  A key aspect of that will be to generate data to train the approximation.  We follow \cite{liu_learning_2023} and use a class of deformation trajectories that are smooth but have peaks and valleys.

We take the total time interval $[0,1]$ (after normalization) and divide it into $M$ intervals $(\Delta t)^m=t^m-t^{m-1}$, $m=1, ...,M$. Where $0 \le t^0<t^1<...<t^M=1$, $\{ (\Delta t)^m \}$ are drawn from a uniform distribution.   We then define
$(F_{ij})^m=(F_{ij})^{m-1}+(v_{ij})^mF_\text{max} \sqrt{(\Delta t)^m}, \  i,j=1,2\le j$
where $(v_{ij})^m \in \{-1,1\}$ follow a Rademacher distribution for each $ij$.  Then, a random path $F_{ij}(t)$ is obtained by cubic Hermite interpolation of $\{ ( t^m,(F_{ij})^m ) \}$.  

A typical trajectory with $M=2$ is shown in Fig.~\ref{fig:randomFS}(a)
and the resulting stress is shown by solid lines in Fig.~\ref{fig:randomFS}(b).  In generating these results, we discretize the overall time interval $[0,1]$ into $N=1500$ time-steps.

\subsubsection{(An)isotropy}

We close this section by studying the slight anisotropy of the typical truss structure shown in Fig.~\ref{fig:truss}(a).  For any imposed deformation trajectory, $F(t)$, we compare the stresses $\sigma(t)$ and $\tilde \sigma(t)$
calculated from $F(t)$ and the rotated trajectory $F(t)R$, respectively, where $R$ denotes rotation.
We define a root squared anisotropy (RSA) as
\begin{equation} \label{eq:rsa}
    \text{RSA}(t)= \frac{  |\sigma(t) - \tilde \sigma(t)| }{  |\sigma(t)| }.
\end{equation}

\begin{figure}
	\begin{center}
        {\includegraphics[width=6in]{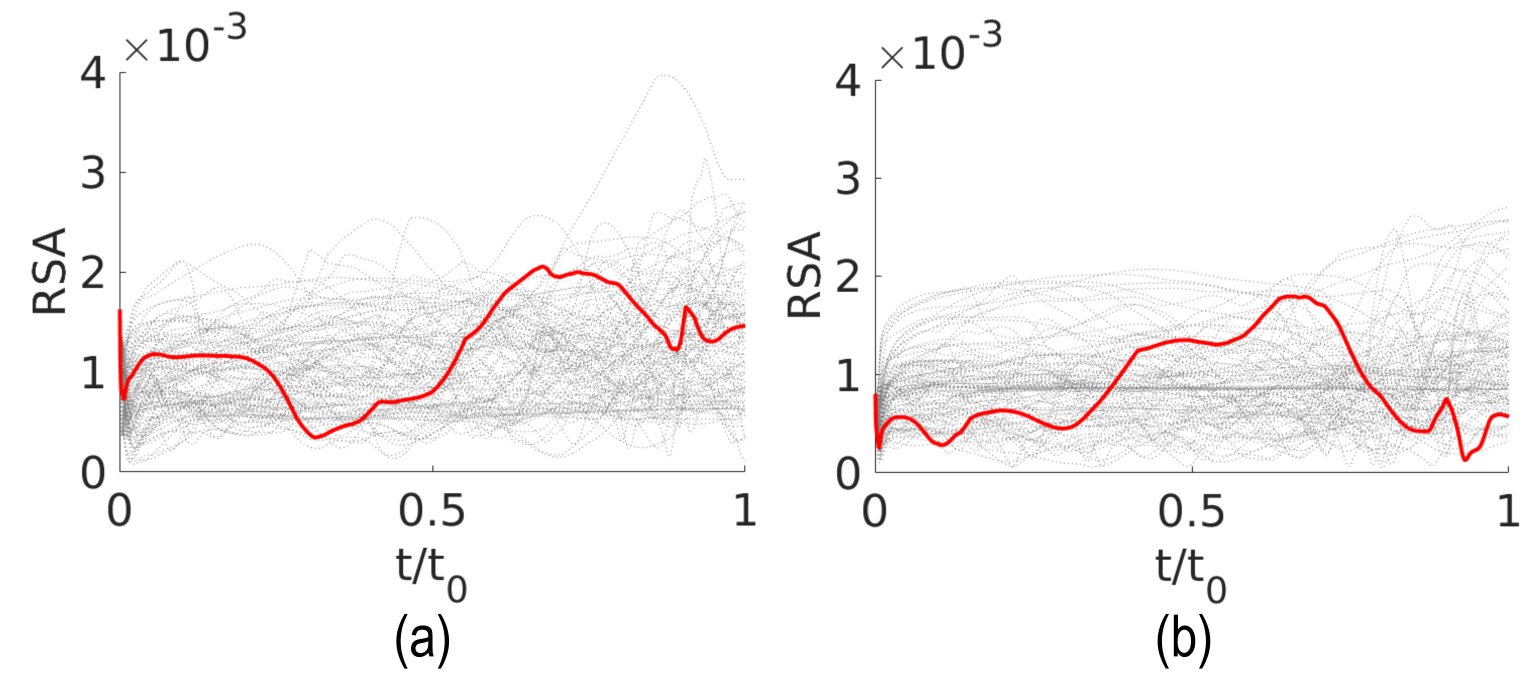}}
		\caption{Root squared anisotropy (RSA), (\ref{eq:rsa}),
			calculated with solutions from 100 realizations of random trajectories
			described in Section \ref{sec_HomoF}. 
			The RSA of one realization is emphasized with red for illustration.
			The rotation angle is (a) $\pi/4$ and (b) $\pi/2$.}
		\label{RSE_anisotropy}
	\end{center}
\end{figure}

Fig.~\ref{RSE_anisotropy} shows the RSA calculated from 100 random trajectories generated as described in Section \ref{sec_HomoF}. 
The angle of rotation is $\pi/4$ in (a) and $\pi/2$ in (b), where the RSE of one realization is emphasized with red for illustration in 
both (a) and (b).
The homogenized Cauchy stress better approximate isotropic response
as the RSE is lower when the structure has more elements and nodes.
For both rotation angles $\pi/4$ and $\pi/2$, the RSE values are in the order
of $10^{-3}$, indicting a good approximation of isotropic response.
Thus the structural material is nominally isotropic.

\section{Recurrent neural operator surrogate}
\label{sec:rno}

In this section, we develop a recurrent neural operator (RNO) approximation for the effective behavior (\ref{eq:effective}) and train it using data generated by repeated solution of the square truss described above using random trajectories.  

\subsection{Recurrent neural operator}

The effective behavior (\ref{eq:effective}) is history dependent, i.e., the current stress depends on the history of the deformation. We follow \cite{liu_learning_2023} and the long tradition in continuum mechanics  \citep[e.g.,][]{rice_inelastic_1971,simo_computational_1998} to represent history using internal or state variables.  Therefore, we seek an approximation of  (\ref{eq:effective}) as 
\begin{equation}
	\begin{cases}
		\sigma(t)=f(F(t),\{ \xi_{\alpha}(t) \}^k_{\alpha=1} ),\\
		\dot{\xi}_i(t)=g_i(F(t),\{ \xi_{\alpha}(t) \}^k_{\alpha=1} )~~ i=1, ..., k
	\end{cases}       
\end{equation}
where $f: \mathbb{R}^{d \times d} \times \mathbb{R}^k \rightarrow \mathbb{R}^{d \times d}_{sym}$
and $g_i: \mathbb{R}^{d \times d} \times \mathbb{R}^k \rightarrow \mathbb{R}$
are (hyper-parametrized) neural networks and the $\{\xi_{\alpha}(t) \}^k_{\alpha=1}$ are internal or state variables.  Unlike classical continuum mechanics, the internal variables are not prescribed {\it a priori}, but are to be determined as a part of the training from data.  The neural networks $f, \{g_i\}$ are also to be trained using data.

We have formulated the RNO as an operator with time being continuous.  In practice, we use a time discretization:
\begin{equation}
	\begin{cases}
		\sigma^n=f(F^n,\{ \xi^n_{\alpha} \}^k_{\alpha=1} ),\\
		\xi_i^{n}={\xi}^{n-1}_i+(\Delta t)_n g_i(F^{n},\{ \xi_{\alpha}^{n-1} \}^k_{\alpha=1} ),~~ i=1, ..., k.
	\end{cases}
    \label{Eq_discretization}
\end{equation}
The time-step ${(\Delta t)_n}$ is not pre-determined or fixed, and in fact does not have to be uniform in $n$\footnote{The time discretization ${(\Delta t)_n}$ is independent of the time intervals ${(\Delta t)^m}$ used to generate random trajectories.}.  Importantly, it is possible that we have one time discretization for the training data (or even multiple time discretizations for different data points),  and an entirely different one for the testing.  In this sense, the RNO is independent of the time discretization.

In what follows, we use neural networks of various depths for $f, \{g_i\}$ with the width of 200 at each layer.  We also consider various numbers of internal variables from 0 to 20.

\subsection{Data and Training}

We generate data $\{ F(t), \sigma(t)\}$ by evaluating the average stress $\sigma(t)$ over the square truss of Fig.~\ref{fig:truss}(a) subjected to a homogenous boundary condition corresponding to 5000 random trajectories generated according to Section \ref{sec_HomoF}.  The data is generated by 1500 time steps over the interval $[0,1]$, and then downsampled to 500 time steps.  We use the results of 3200 trajectories for training, 800 for validation and  1000 for testing the neural approximation (\ref{Eq_discretization}).  We consider the following error (loss) function in training and testing:
\begin{equation}
	\text{error}=\frac{1}{S} \sum^{S}_{s=1}
	\left(
	\frac{\int^T_0 | \sigma_s^\text{RNO}(t) - \sigma_s^\text{truth}(t) |^2 dt
	}{\int_0^T | \sigma_s^\text{truth}(t) |^2 dt }
	\right)^{1/2}
	\label{Eq_err}
\end{equation}
where $s$ indexes the $S$ trajectories.  We use min-max normalization for all data.

We use an Adam optimizer for first-order gradient-based optimization of stochastic objective functions \citep{kingma_adam_2014}.  The learning rate is 0.001 and batch size is 32.

\subsection{Results}

\begin{figure}[t]
	\begin{center}
		{\includegraphics[width=5.5in]{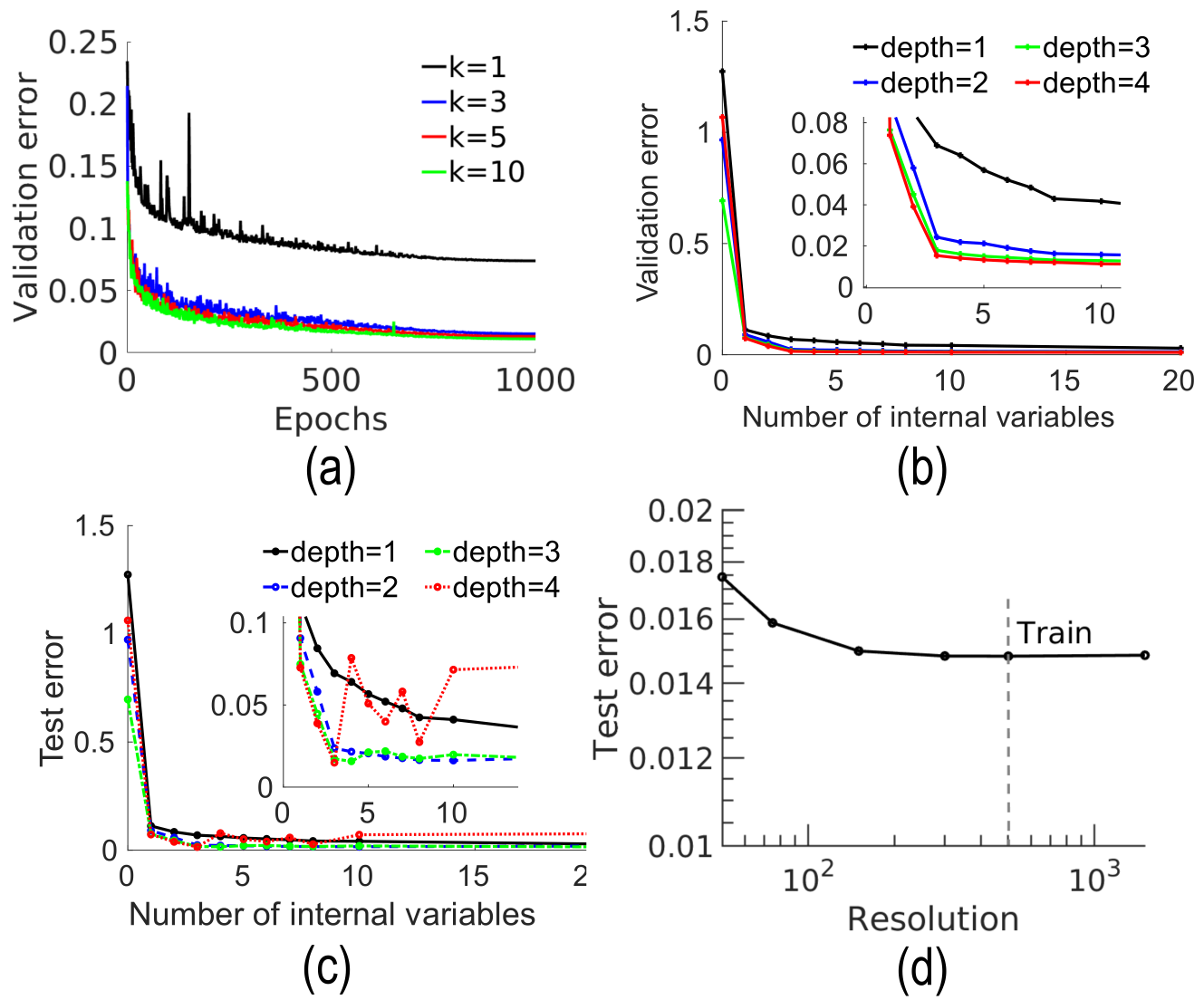}}
		\caption{Recurrent neural operator approximation of the effective behavior of a square truss.
			(a,b) Validation errors vs.\ (a) epochs (depth=4, various number of internal variables) and (b) number of internal variables (various  depths, epochs = 1000).
			(c,d) Test error vs.\ (c) number of internal variables (various  depths) and (d) resolution of input data
			(depth=4 layers with 3 internal variables).}
		\label{fig:TrainValidTestErr_RNO1}
	\end{center}
\end{figure}

The results of the training and the RNO approximation to the effective behavior of the square truss are shown in Fig.~\ref{fig:TrainValidTestErr_RNO1}.  The test and validation errors for various epochs, depths and number of internal variables are shown in Fig.~\ref{fig:TrainValidTestErr_RNO1}(a,b).  We observe that the error decreases with increasing training epochs and layer depths, saturating at about 1000 epochs of training with 4 layers.  The errors drop dramatically initially with increasing number of internal variables, but then saturate with an elbow at three internal variables; it fluctuates a little beyond that with no discernible trend. The training and validation errors
are about 1.5\% with about 1000 epochs of training, 4 layers and 3 internal variables.  Fig.~\ref{fig:TrainValidTestErr_RNO1}(c) shows that the test error as a function of internal variables.  We again see that the test errors drop dramatically initially but saturate at 2\% with an elbow at 3 internal variables.
For future use, we label the RNO with four layers and three internal variables trained with the 4000 random trajectories RNO$_1$.

We conclude that an RNO with three internal variables is able to approximate the effective behavior of the square random truss to a high degree of accuracy.  
We make three observations.  First, the elbow behavior of the error with the number of internal variables, and good approximation with a small number of internal variables is consistent with the experience in crystal plasticity \citep{liu_learning_2023}, linear viscoelasticity \citep{bhattacharya_learning_2023} and reactive flow through porous media \citep{karimi_learning-based_2023}.   Second, we recall that \cite{liu_learning_2023} found that two internal variables were sufficient to represent the effective behavior of elastic-viscoplastic polycrystals in two dimensions.  In their continuum crystal plasticity, the slip systems and consequently the overall plastic deformation is isochoric and therefore two internal variables were sufficient to represent the inelastic deformation in two dimensions.  The plastic deformation of our truss structure is not necessarily isochoric, and therefore isotropy requires three internal variables.  This is what we observe here.  Finally, the fact that we can represent the overall behavior of square random truss with three internal variables is a remarkable reduction in the complexity of the underlying model.  Our truss had 113 links and 20 internal nodes, and thus 266 internal degrees of freedom.  Therefore, we may view our result as an exercise in model reduction: the trained RNO with three internal degrees of freedom is able to emulate the behavior of a system with 266 internal degrees of freedom.

We conclude by demonstrating time discretization invariance.  We return to our data generated by 1500 time steps and down sample it to various coarser time discretizations.  We compare the results with that of an RNO trained at 500 time step discretization.  We see in Fig.~\ref{fig:TrainValidTestErr_RNO1}(d) that the error is independent of the discretization as anticipated by the architecture.

\section{Macroscale simulation of a plate}
\label{sec:FEM}

\begin{figure}[t]
	\begin{center}
		\includegraphics[width=6.5in]{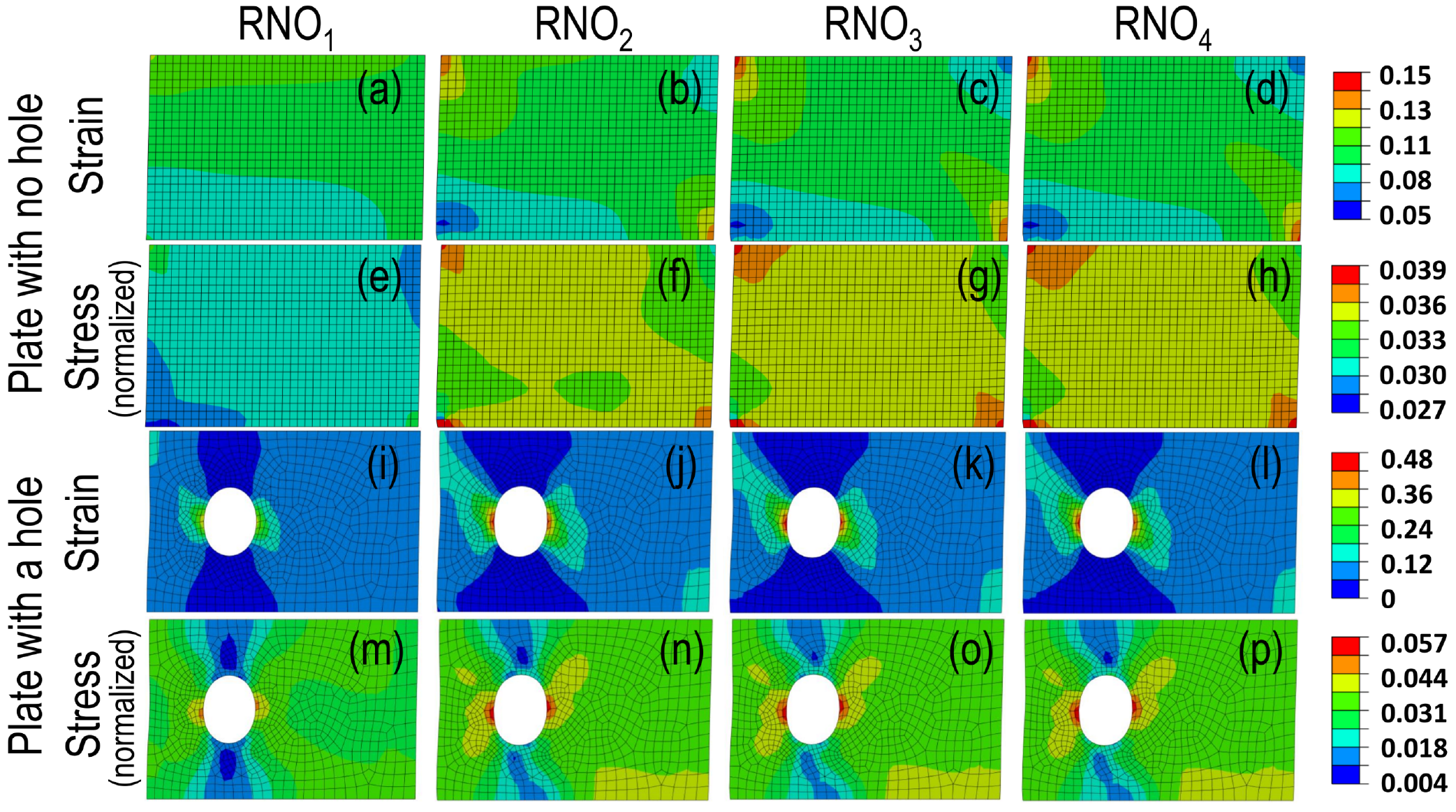}
		\caption{The results of finite element calculations on the full plate and the plate with a hole for variously trained RNOs.  (a-h) Full plate.  (i-p) Plate with a hole. 	
			(a-d, i-l) Contours of maximum in-plane principle strain and (e-h, m-p) contours of normalized Mises stress of the plate.  The calculations use (a,e,i,m) RNO$_1$, (b,f,j,n) RNO$_2$, (c,g,k,o) RNO$_3$ and (d,h,l,p) RNO$_4$
			trained in Sections \ref{sec:rno} and \ref{sec:transferLearning} as the material model.}
		\label{fig:plate_FEM}
	\end{center}
\end{figure}

In this section, we use RNO$_1$ trained in the previous section to study the macroscale simulation of a plate with and without a hole.  The goal is two-fold.  The first is that one can indeed use such RNOs in macroscale simulations and the second is to understand the {\it a posteriori} error in using the RNO as a surrogate for the microscale random truss in a multiscale simulation.

We implement RNO$_1$ as a material model (VUMAT) in the finite element software ABAQUS \citep{dassault_systemes_simulia_corp_abaquscae_2020}.  We consider a plate of length  $5 \times 3 l_0$ (for some $l_0 >>1$  in non-dimensional unit introduced earlier; since our theory has no internal length scale, our results presented in stresses and deformation gradients is independent of $l_0$).  We consider a full plate and a plate with a hole of radius $0.5 l_0$ centered at $(1.5l_0,1.5l_0)$ from the bottom left corner.   
The full plate is discretized with 888 nodes and 828 bilinear plane stress quadrilateral elements  with reduced integration and hourglass control.  The plate with a hole has 884 nodes and 821 elements.  The calculation is carried out with 500 time steps.

We fix the left corner of the plate, and constrain the vertical displacement and the shear stress along the bottom surface to be zero (so it can slide horizontally freely).  We apply a vertical displacement $\hat{u}(t)$ that is initially shifted cosinusoidal starting at 0 with increasing tangent from 0, then linear in time till it reaches a strain of 0.1 ($\hat{u}(1) = 0.3 l_0$) to the top surface. 
Similarly, the shear stress along the top surface is constrained to be zero.
The lateral sides are traction-free.

The results are shown in Fig.~\ref{fig:plate_FEM}(a,e) for the full plate.  We observe a stress concentration at the left corner in the full plate due to the constrain there. The results are shown in  Fig.~\ref{fig:plate_FEM} (i,m) for the plate with a hole.   We see stress concentration around the hole.  These results show that one can indeed use the RNO in macroscale calculations consistent with previous works \citep{liu_learning_2023,karimi_learning-based_2023}.

We now turn to understanding the error of the RNO trained on random trajectories in the actual trajectories encountered in these calculations.  We extract 828 deformation gradient trajectories from the 828 Gauss points in the full plate.  For these trajectories, we compute the stress using both the square truss model (Section \ref{sec:truss}) and RNO$_1$, and use these to compute the error (\ref{Eq_err}).  We find an error of 0.045.  The corresponding error using the 821 deformation gradient trajectories show an even larger error of 0.06.  While these are still reasonable, they are almost 3-4 times the training and test errors noted earlier (Section \ref{sec:rno}).   

Finally, we note for further comment that there is another form of error that this analysis does not account for.  The trajectories were generated using RNO$_1$ as the material model in the finite element calculations.  A more accurate model would have different trajectories.

\section{Iterated learning}
\label{sec:transferLearning}

We saw in the previous section that the trajectories involved in the FEM lead to significant errors in the RNO trained with random trajectories.   One possible source of this error is that the random trajectories are not representative of the trajectories encountered in the FEM.  Indeed, observe in Fig.~\ref{fig:plate_FEM}(i) that the strains (near the hole) are in fact quite large.  So we study iterative learning in this section where we repeatedly update the training dataset from finite element calculations.

\subsection{Iterated learning}

We start with the 5000 random deformation gradient trajectories described in Section \ref{sec_HomoF} and the corresponding stress histories computed using the square truss and label this dataset ${\mathcal D}_1$.  We use 3200 and 800 trajectories from this dataset to train and validate the RNO, and call the trained model RNO$_1$.  We then generate dataset ${\mathcal D}_1^\text{FEM}$ by using RNO$_1$ in a finite element analysis, extract the deformation gradient trajectories from the finite element calculation and compute the stress histories for these trajectories using the square truss.  

We then generate a second dataset ${\mathcal D}_2$ from a combination of ${\mathcal D}_1$ and ${\mathcal D}_1^\text{FEM}$.  We use this in transfer learning: we start with RNO$_1$ and use the dataset ${\mathcal D}_2$ to update the hyper-parameters.  We call this (transfer) trained model RNO$_2$.  We then use this to study the same finite element analysis as above and generate dataset ${\mathcal D}_2^\text{FEM}$.  We iterate in this manner, in each step creating a new dataset based on the immediate past finite element analysis and all previous datasets.  The details are provided in Table \ref{tab:data}.

\begin{table}
\caption{Iterative generation of datasets and models \label{tab:data}}
\vspace{0.1in}
\centering
\begin{tabular}{|p{0.3in}|p{1.1in}|p{1.1in}|p{1.2in}|p{0.8in}|p{0.8in}|}
\hline 
& Train $( {\mathcal D}_i^\text{train}) $ & Validate $({\mathcal D}_i^\text{val} )$ & Test $({\mathcal D}_i^\text{test} )$ & Model & Generate\\
\hline 
${\mathcal D}_1$ & 3200 random & 800 random & 1000 random & RNO$_1$ & ${\mathcal D}_1^\text{FEM}$\\
\hline
${\mathcal D}_2$ & 480 random, 480 ${\mathcal D}_1^\text{FEM}$ & 124 random, 124 ${\mathcal D}_1^\text{FEM}$ & remainder ${\mathcal D}_1^\text{FEM}$ & RNO$_2$ & ${\mathcal D}_2^\text{FEM}$\\
\hline
\multicolumn{6} {|c|}{\dots}\\
\hline
${\mathcal D}_n$ & 480 ${\mathcal D}_{n-1}^\text{FEM}$, $\left\{ \frac{480}{n-1} {\mathcal D}_{i}^\text{FEM}\right\}_{ i = 1}^{n-2}$, $\frac{480}{n-1} {\mathcal D}_1^\text{train}$ 
 & 124 ${\mathcal D}_{n-1}^\text{FEM}$, $\left\{ \frac{124}{n-1} {\mathcal D}_{i}^\text{FEM}\right\}_{ i = 1}^{n-2}$, $\frac{124}{n-1} {\mathcal D}_1^\text{val}$  & remainder ${\mathcal D}_{n-1}^\text{FEM}$ & RNO$_n$ & ${\mathcal D}_n^\text{FEM}$\\
\hline
\end{tabular}
\end{table}

\begin{figure}[t]
	\begin{center}
		{\includegraphics[width=5.5in]{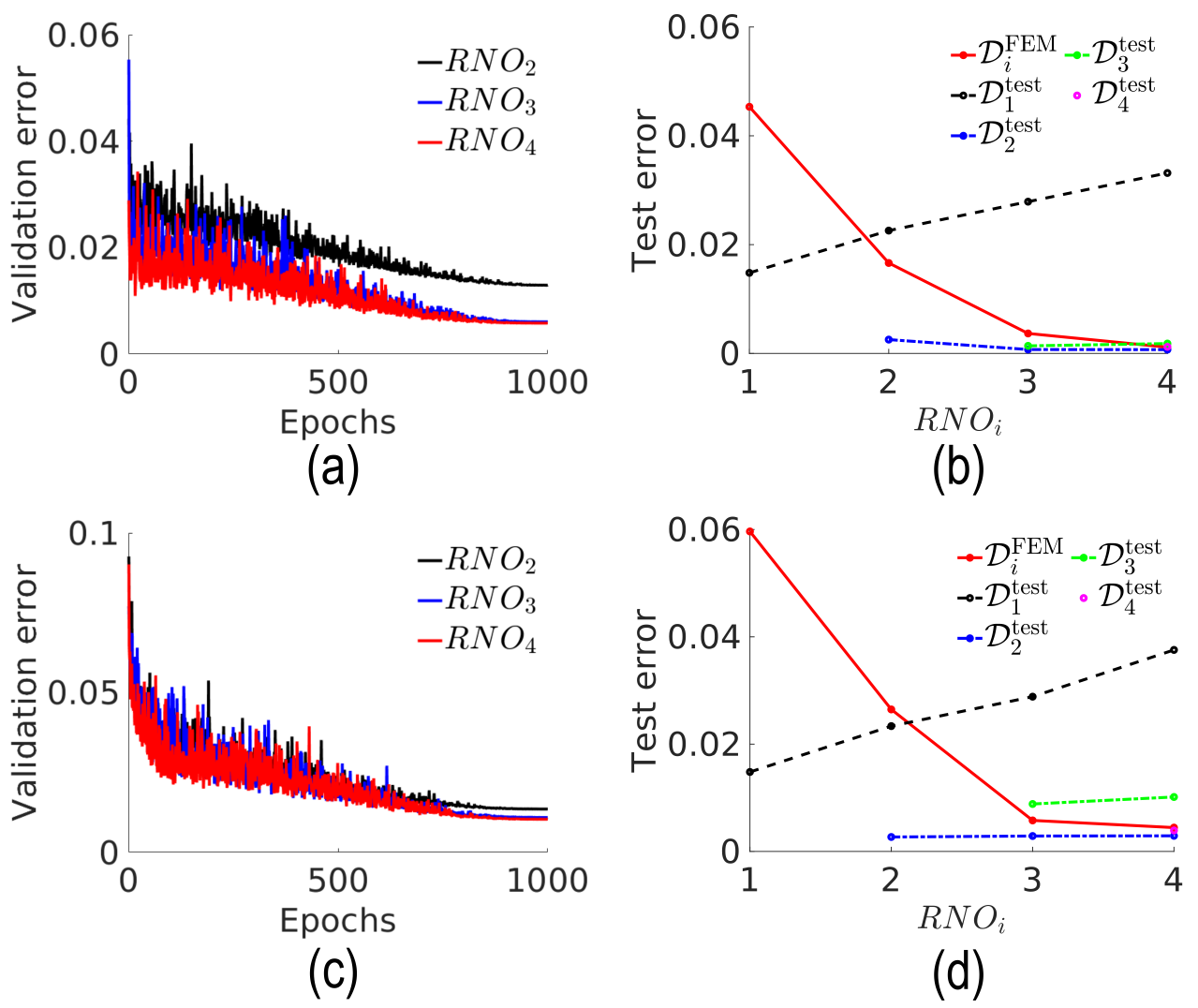}}
		\caption{The evolution of errors during the iterative training process. 
			 (a,b) Full plate, (c,d) Plate with a hole.
			 (a,c) Validation errors during the successive training, 
			 (b,d) Test error of the trained models against various datasets.}
		\label{fig:iterative_errors}
	\end{center}
\end{figure}

We follow this process independently for the full plate, and the finite element analysis of the plate with a hole.  Therefore we have two sets of data and models.

\subsection{Errors and Stresses}

The training and validation as well as various test errors for the successive models are shown in Fig.~\ref{fig:iterative_errors} for the full plate (Fig.~\ref{fig:iterative_errors}(a,b)) and the plate with a hole (Fig.~\ref{fig:iterative_errors}(c,d)).    We see Fig.~\ref{fig:iterative_errors}(a,c) that the initial training and validation error for models $n\ge 2$ starts at a value that is smaller than the corresponding error for RNO$_1$
in Fig.~\ref{fig:TrainValidTestErr_RNO1}(a).  It reduces on (transfer) training with the new dataset and reaches a value comparable to the final training error of RNO$_1$. 

The test errors are shown in Fig.~\ref{fig:iterative_errors}(b) for the full plate and Fig.~\ref{fig:iterative_errors}(d) for the plate with a hole.  We focus on the error of model RNO$_n$ on the dataset ${\mathcal D}_n^\text{FEM}$ that is generated from the finite element analysis using the same model (shown in the red solid line).  As already noted in the previous section, this is about 4.5\% for the full plate and 6\% for the plate with a hole for RNO$_1$.  However, it drops with each successive model and saturates at values significantly less than 1\% by RNO$_4$.  In other words, the iterative training procedure is able to produce a highly accurate finite element simulation.

Fig.~\ref{fig:plate_FEM} shows that the strain and stress distribution from the finite element analysis also evolve with each successive model, but stabilizes between RNO$_3$ and RNO$_4$.  The stress concentrations at the corners of the plate and the vicinity of the hole become sharper and well defined.  Since the initial model RNO$_1$ was trained on data with limited strains, it is unable to resolve these high strain regions.  However, the later models are able to learn these relevant strains as a result of the iterative process.

\begin{figure}
	\centering
	{\includegraphics[width=5.5in]{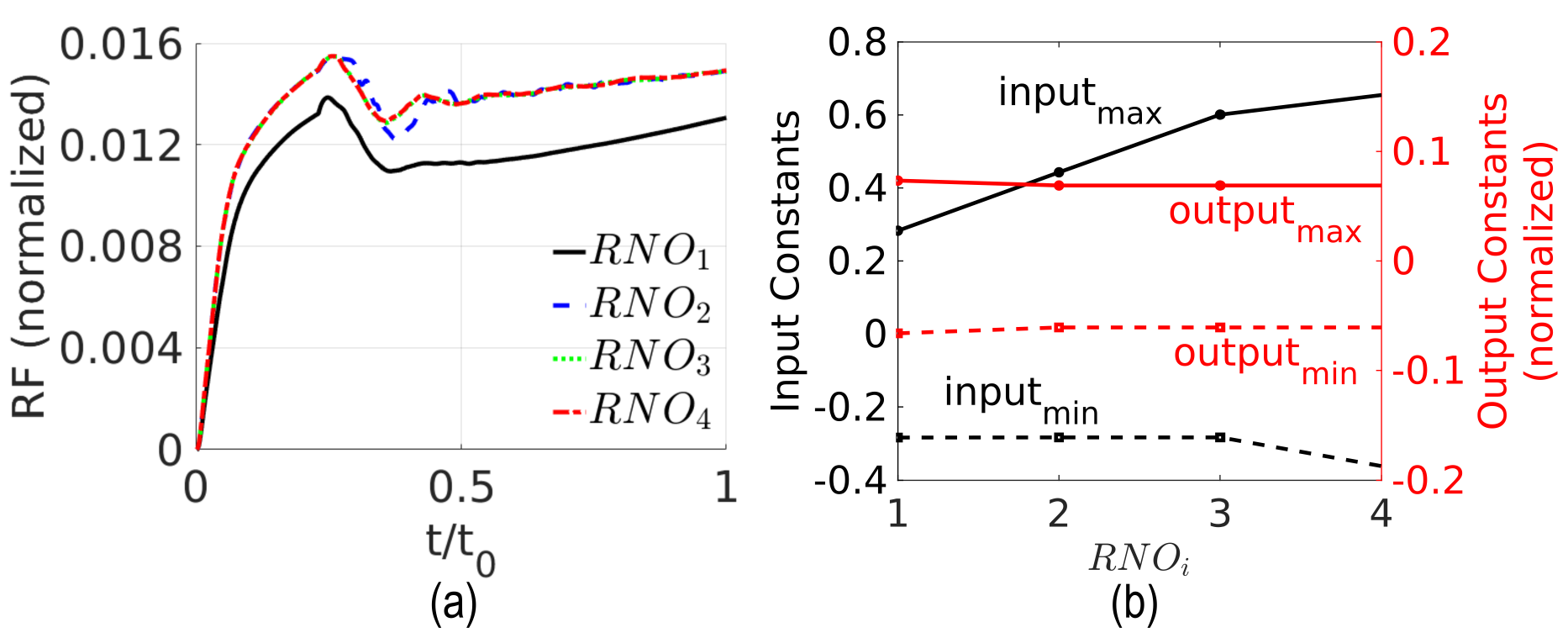}}
	\caption{Analysis of iterative learning for a plate with a hole. (a) The total vertical reaction force (RF) vs. time. (b) The constants used for the min-max normalization of the data during the training process.} 
	\label{fig:plate_hole}
\end{figure}

We examine this further for the plate with a hole in Fig.~\ref{fig:plate_hole}.  Fig.~\ref{fig:plate_hole}(a) shows the total applied force vs.\ time.  We see that there is a slight drop in the force when the stress material first yields near the hole, but it then recovers and stays steady.   The level of force increases with successive models, but the features remain similar.  This shows that all the models are robust enough to handle load drops without becoming unstable.  The results converge by RNO$_4$.    Fig.~\ref{fig:plate_hole}(b) shows the constants used for the min-max normalization of the data during the training process, and provides an insight into the range of the datasets.  We observe that the constants for the inputs span a larger region with every successive training, but the constants for the outputs do not change.  This shows that the range of strains encountered in each successive dataset increases, but the range of stresses does not.  This is related to the formation of the stress concentrations.

\subsection{Data and internal variables}

\begin{figure}[t]
	\begin{center}
		\includegraphics[width=5.5in]{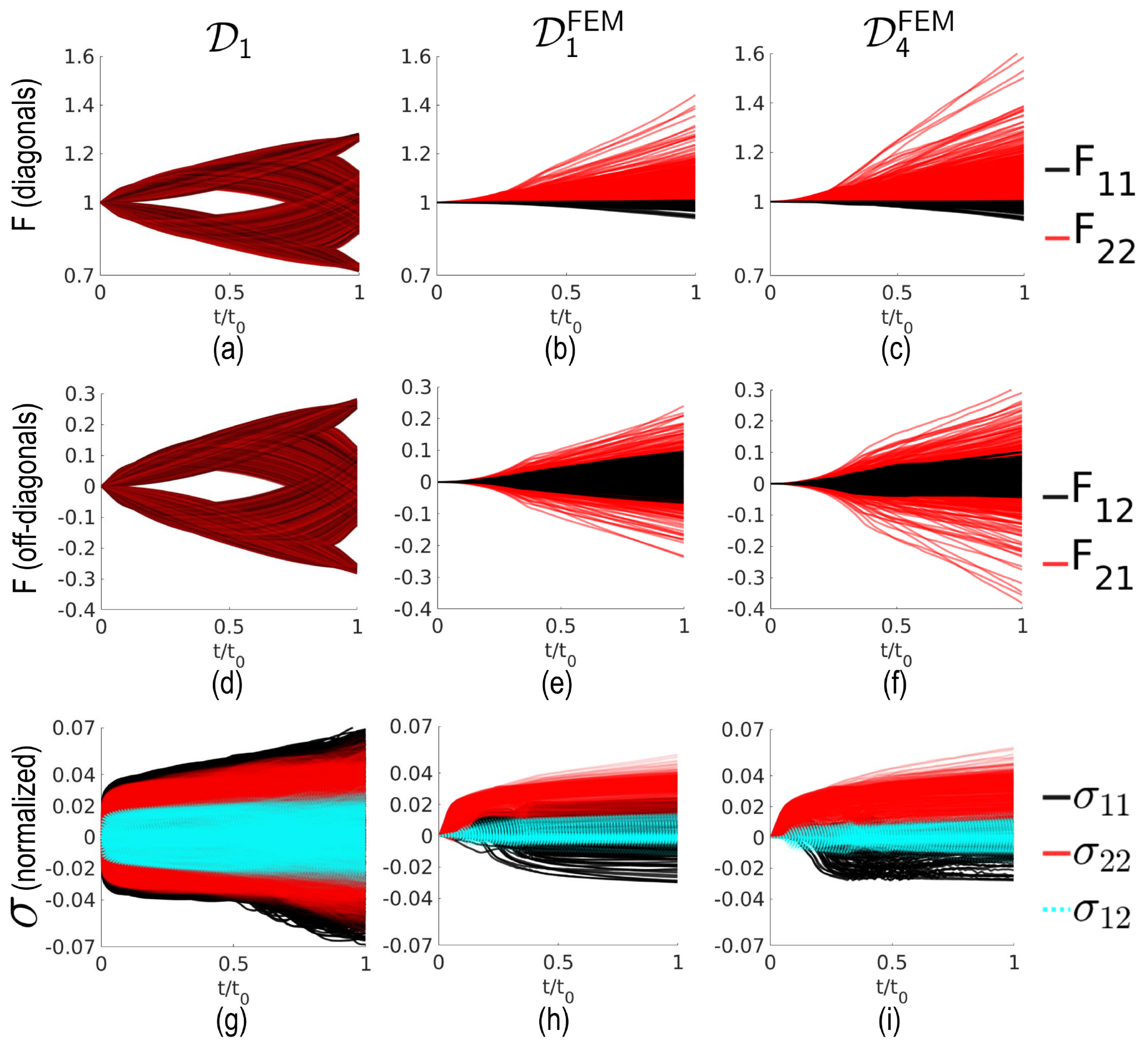}
		\caption{Sample space taken by the trajectories of deformation gradient
			$F(t)$ in (a-f) and normalized stress $\sigma(t)$ in (g-i)
			for the plate with a hole. 
			(a,d,g) dataset ${\mathcal D}_1$,
			(b,e,h) dataset ${\mathcal D}_{1}^\text{FEM}$,
			(c,f,i) dataset ${\mathcal D}_{4}^\text{FEM}$.}
		\label{fig:data_range}
	\end{center}
\end{figure}

We examine the change of the range of data and internal variables as we proceed through the iterative training to gain an insight into the reason why iterated training leads to greater accuracy.

Fig.~\ref{fig:data_range} shows the trajectories associated with the datasets ${\mathcal D}_1$ (the original dataset), ${\mathcal D}_1^\text{FEM}$ (the trajectories generated by the finite element method with RNO$_1$), and ${\mathcal D}_4^\text{FEM}$  (the trajectories generated by the finite element method with RNO$_4$).  We see that the trajectories connected with the original dataset ${\mathcal D}_1$ are broadly distributed in all components of stress and strain.  However, the trajectories generated by the finite element method involve large values of $F_{22}$ and $F_{21}$, but smaller values of $F_{11}$ and $F_{12}$.  Importantly, some trajectories in ${\mathcal D}_{1,4}^\text{FEM}$ involve larger values of $F_{22}$ than those in ${\mathcal D}_1$.  Similarly, we see large values of $\sigma_{22}$, but relatively small values of $\sigma_{11}$ and $\sigma_{12}$, in ${\mathcal D}_{1,4}^\text{FEM}$.  Importantly, we note a very significant change in trajectories at the first FEM iteration, but relatively change from the first to the fourth iteration.

\begin{figure}
	\begin{center}
		\includegraphics[width=6.5in]{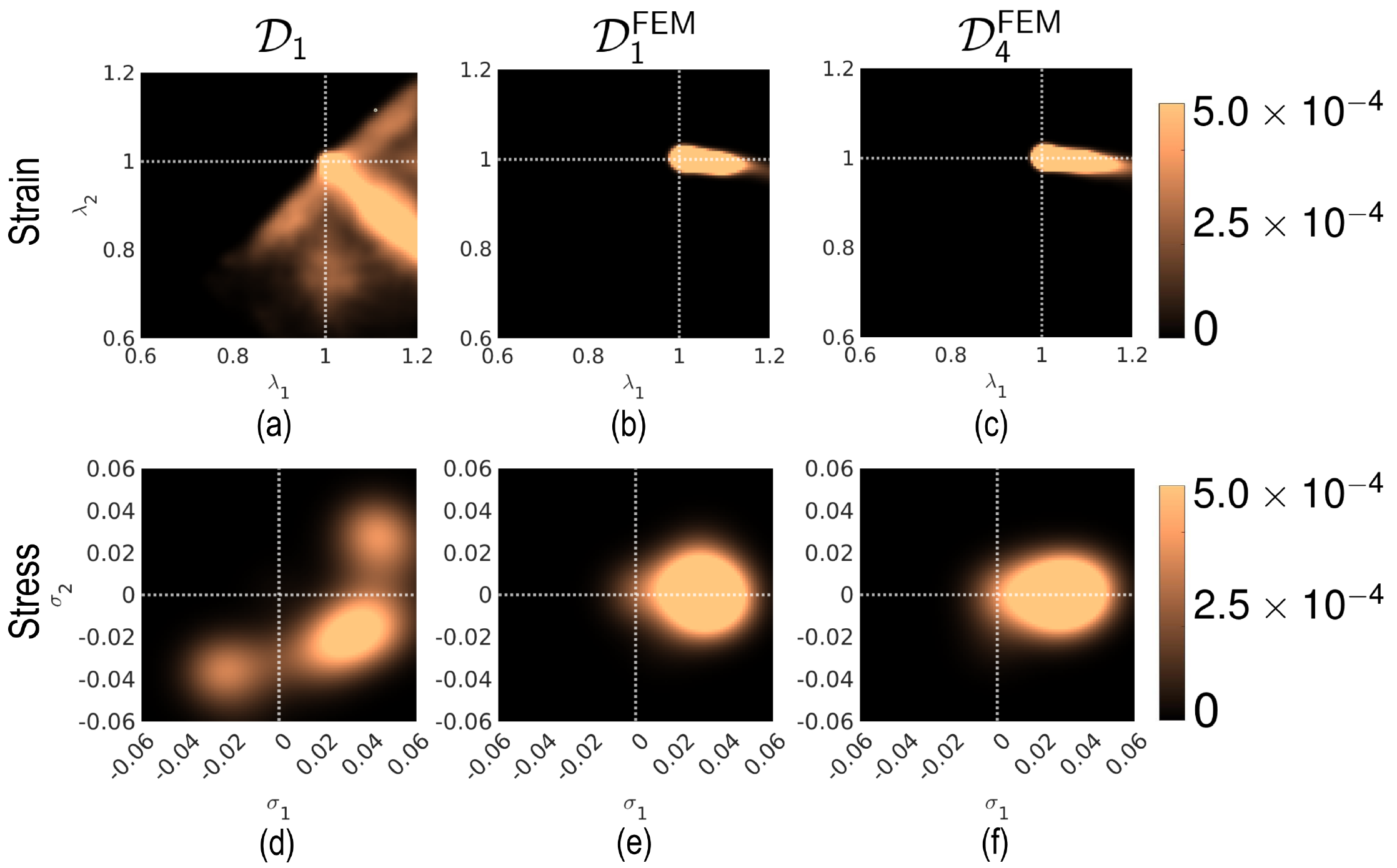}
		\caption{Probability density of principal strains $(\lambda_1,\lambda_2)$
			and stresses $(\sigma_1,\sigma_2)$ of various datasets
			for the plate with a hole. 
			(a,d) dataset ${\mathcal D}_1$,
			(b,e) dataset ${\mathcal D}_{1}^\text{FEM}$,
			(c,f) dataset ${\mathcal D}_{4}^\text{FEM}$.
			The frequency before normalization is counted with a normalized time step size 0.01.}
		\label{fig:data_principal}
	\end{center}
\end{figure}

Fig.~\ref{fig:data_principal} shows this change in a different way.  For each trajectory in a dataset, and at every 3$^\text{rd}$ time-step, we find the principal strains (singular values of $F$) and principal stresses (eigenvalues of $\sigma$).   Fig.~\ref{fig:data_principal} shows the density plots of these principal stresses and strains for the datasets ${\mathcal D}_1$, ${\mathcal D}_1^\text{FEM}$ and  ${\mathcal D}_4^\text{FEM}$.  Again, we see that the principal strains and stresses are broadly distributed in ${\mathcal D}_1$, but quickly change to concentrated towards uniaxial strain and stress in ${\mathcal D}_4^\text{FEM}$.

\begin{figure}[t]
	\begin{center}
		\includegraphics[width=6in]{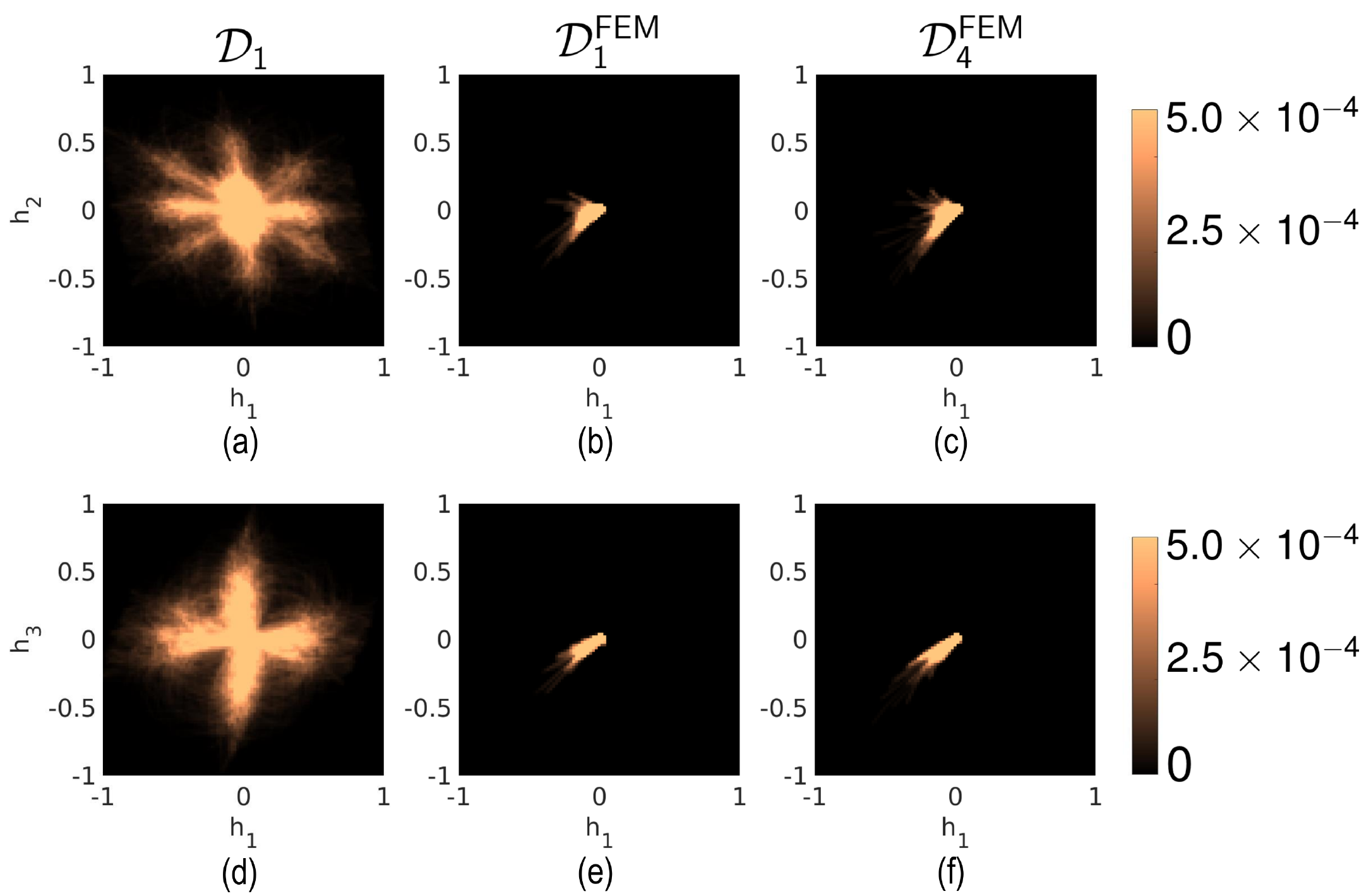}
		\caption{Probability density of internal variables $(h_1,h_2,h_3)$
	        of various datasets for the plate with a hole. 
			(a,d) dataset ${\mathcal D}_1$,
			(b,e) dataset ${\mathcal D}_{1}^\text{FEM}$,
			(c,f) dataset ${\mathcal D}_{4}^\text{FEM}$.
			The frequency before normalization is counted with a normalized time step size 0.01.}
		\label{fig:param_range}
	\end{center}
\end{figure}

The range of the values of the internal variables accessed by corresponding RNOs in the datasets is shown in Fig.~\ref{fig:param_range}.  We again see a change from a broad range to a concentration.

These observations lead us to conclude the following about iterated training.
The initial model is trained on random trajectories, and thus leads to a model that is reasonably accurate giving it a robustness against all possible loadings.  Thus, the initial model is sufficiently accurate for finite element calculations to lead to reasonably accurate trajectories.  Such trajectories increase the range in certain directions, but more importantly are concentrated in limited areas.  Oversampling these trajectories (by using a mix of random and FEM derived trajectories) lead to greater accuracy.

\subsection{Transferability of iterated learning}

\begin{table}
\centering
\caption{Transferability of iterated learning: Errors against various datasets \label{tab:transfer}}
\vspace{0.1in}
\begin{tabular}{|c|c|c|c|c|c|c|c|c|}
\hline
Model & VT$^\text{NH}$ & VC$^\text{NH}$ & HT$^\text{NH}$ & SS$^\text{NH}$
& VT$^\text{H}$ & VC$^\text{H}$ & HT$^\text{H}$ & SS$^\text{H}$ \\
\hline
RNO$_1$ (black)  
&0.045 & 0.073 & 0.068 & 0.083       
&0.060 & 0.080 & 0.085 & 0.080\\     
\hline
RNO$_{VT4}^\text{NH}$ (blue) 
&0.001 & 0.070 & 0.043 & 0.044        
&0.022 & 0.077 & 0.048 & 0.046\\      
\hline
RNO$_{VT4}^\text{H}$ (red)
&0.002 & 0.039 & 0.038 & 0.044        
&0.004 & 0.051 & 0.041 & 0.039 \\      
\hline
\multicolumn{9}{|c|}{\parbox{4.5in}{ VT$^\text{NH}$: Vertical tension with no hole (10\% strain), \\
VC$^\text{NH}$: Vertical compression with no hole (10\% strain), \\
HT$^\text{NH}$: Horizontal tension with no hole (10\% strain), \\
SS$^\text{NH}$: Simple shear with no hole (10\% strain), \\
VT$^\text{H}$: Vertical tension with hole (10\% strain), \\
VC$^\text{H}$: Vertical compression with hole (10\% strain), \\
HT$^\text{H}$: Horizontal tension with hole (10\% strain), \\
SS$^\text{H}$: Simple shear with hole (10\% strain)}} \\
\hline
\end{tabular}
\end{table}

\begin{figure}[t]
	\begin{center}
		\includegraphics[width=4in]{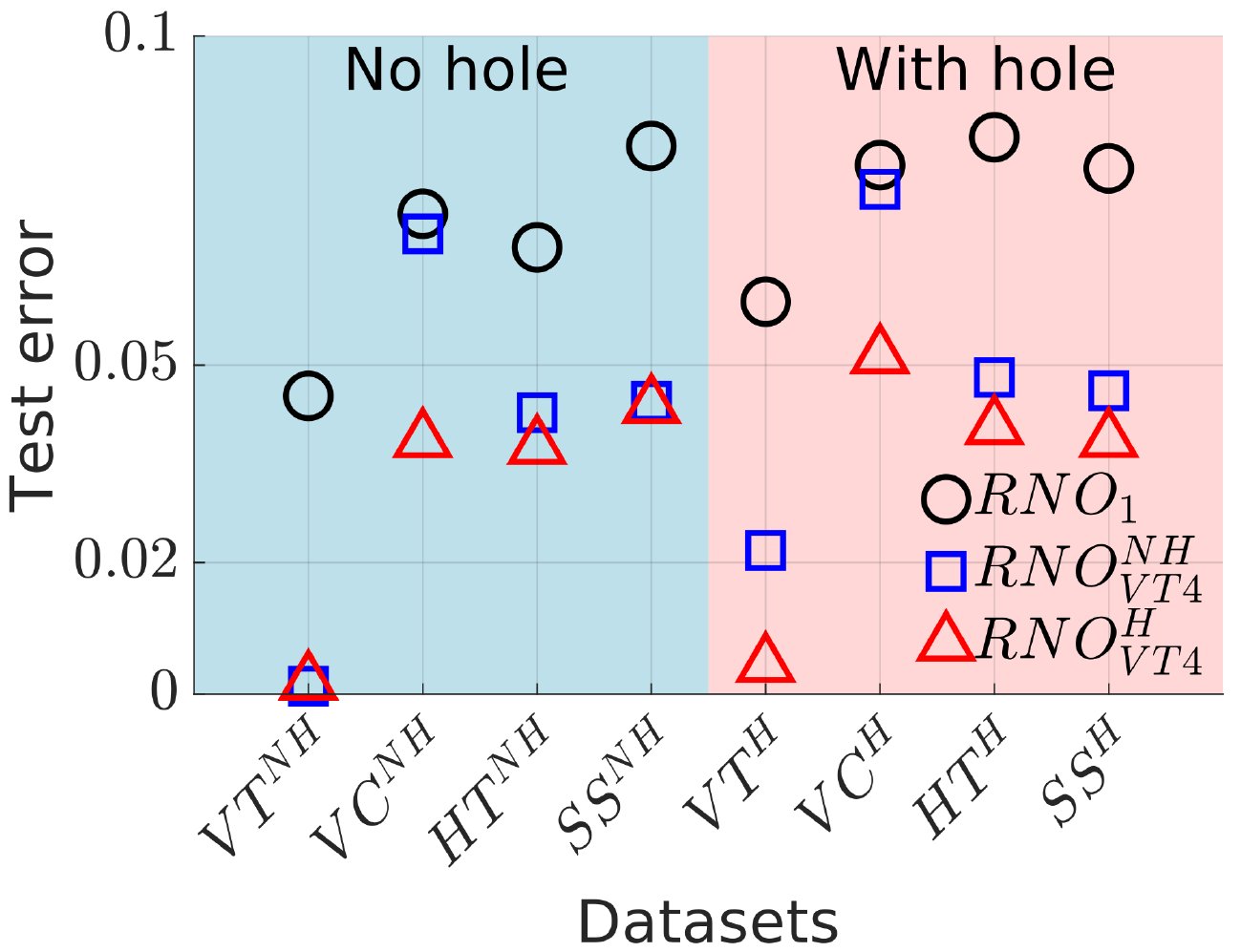}
		\caption{Transferability of iterated learning: Errors against various datasets.}
		\label{fig:transferability}
	\end{center}
\end{figure}

The previous section showed that iterated learning is an effective approach to reducing the error of the RNO model in finite element calculations.  We now examine transferability of the iteratively trained RNO, i.e., the performance of the RNO trained iteratively on one calculation on other calculations.

Recall the two sets of iteratively trained RNO: RNO$_i^\text{NH}$, $i=1,\dots 4$ under vertical tension on the specimen with no hole and RNO$_i^\text{H}$, $i=1,\dots 4$ under vertical tension on the specimen with a hole.
We use these models for the finite element analysis of eight distinct problems, the original vertical tension, vertical compression, horizontal tension and simple shear for two specimens, one with no hole and one with a hole.  We then take the trajectories generated in these studies and compute the average error of the RNO on these trajectories.  The results are shown in Table \ref{tab:transfer}
and plotted in Fig.~(\ref{fig:transferability}).

We make the following observations.
\begin{itemize}
\item The iterated RNO performs best on the problem on which it is trained.  This is true for both families of RNOs iteratively trained on problems with and without a hole.
\item RNO$_4$ has smaller error compared to RNO$_1$ in all calculations despite the fact that it is only trained iteratively on vertical tension.  This is true for both families RNOs iteratively trained on problems with and without a hole.  Thus, iteratively training an RNO on one problem leads to a smaller error in all calculations.
\item  RNO$_4^\text{H}$ performs better on a specimen with no hole, compared to RNO$_4^\text{NH}$ on problems on a specimen with a hole. 
Also, except $\rm VT^{NH}$, RNO$_4^\text{H}$ gives smaller error than RNO$_4^\text{NH}$ for all other cases.
In other words, training on a specimen with complicated deformation fields leads to a more versatile approximation.
\item HT and SS have better transferability from $\rm RNO_4$ of VT compared with VC.

\end{itemize}

In summary, iterated training leads to a better surrogate not only for the problem on which it is trained but for other problems as well.  Further, iterated training on complex geometries lead to better approximations.

We conclude by making two more observations. 
First, horizontal tension behaves similarly to vertical tension.  This is a result of the relative isotropy of the truss under consideration.
Second, vertical compression leads to the largest error.  This reflects the fact that a truss with elements undergoing finite deformation has tension-compression asymmetry.

\section{Conclusions}
\label{sec:conc}

We have examined the ability to represent the overall behavior of a random architected material using a recurrent neural operator (RNO), and the data necessary to do so.  We show that in two dimensions, a RNO with three internal variables is able to accurately describe the overall elastic-plastic behavior.  The plastic deformation of an architected material  is not isochoric, and therefore the plastic strain spans three dimensions in a two dimensional space setting.  This is consistent with the minimum required number of internal variables.  This ability of the RNO to represent history-dependent behavior is consistent with previous works in other phenomena \citep{liu_learning_2023,bhattacharya_learning_2023,karimi_learning-based_2023}.

The core issue we examine is the data necessary for such neural network based surrogate models.  Following previous work, we train the RNO on data generated from a rich but arbitrary class of trajectories to find good accuracy.  We then use the trained RNO in a finite element analysis of a macroscopic problem, and find that the accuracy -- while still reasonable -- decreases for the trajectories encountered in the finite element analysis.

The main contribution of the paper is to introduce an iterative approach to training such a surrogate.  In this approach, we first use a rich arbitrary class of trajectories  to train an initial model, and then use this model in a finite element analysis.  We then create a new dataset from a combination of trajectories from the initial rich arbitrary class and the finite element analysis.  We use this dataset for transfer learning from the previously trained surrogate: we initialize with the previously trained surrogate and update the hyper-parameters based on the new combined dataset.  We then use this updated surrogate in a finite element analysis and iterate.

We find that the approach converges rapidly, after four iterations with the error stabilizing to a low value on the trajectories associated with finite element analysis.  Further, the converged surrogate is transferable to a degree: it shows much higher accuracy on macroscopic problems other than the one used in iteration compared to the original model trained on the rich arbitrary paths.  This is especially the case when the macroscopic problem used in the iteration has stress raisers like holes and corners.

A key reason for the decrease in error from the initial trained model to the converged one is due to the difference between trajectories that are used in the initial class and those that arise in the finite element analysis.   First, the initial class was chosen to be span a large region of strain space and the principal strains are broadly distributed.  However, the trajectories that arise in the finite element analysis are much more concentrated on a region with unequal principal strains (high shear).    Second, the trajectories that arise in the finite element analysis explore larger strains in selected regions.   Finally, even the first finite element analysis is quite good in identifying the right class of trajectories.

We believe that this work motivates the need for a systematic examination of the error in the use of neural network surrogates in history-dependent multiscale models, and the relationship between this error and the data.  Specifically, it motivates the need to develop rigorous methods to classify and sample trajectories, and to understand the relation between this sampling and neural network error.

\section*{Acknowledgements}  We gratefully acknowledge the financial support of the Army Research Office through Grant Number W911NF-22-1-0269. The simulations reported here were conducted on the Resnick High Performance Computing Cluster at the California Institute of Technology.

\section*{Data and code}  The data and the codes are available on request.

\renewcommand\thesection{Appendix \Alph{section}}
\setcounter{section}{0}

\section{Elastic-Plastic Evolution}
\label{appendix_numerical}

We consider a time discretization with time step $\Delta t$.  At the $k^\text{th}$ time step, we are given the nodal positions  $\{ y_n^{k-1}\}$,  the linkwise plastic strains $\{ (\epsilon_l^\text{pl})^{k-1} \}$ and the linkwise accumulated plastic work $\{ \alpha_l^{k-1} \}$.  We seek to find the updated values
$$
y_n^k = y_n^{k-1} + \Delta y_n, \quad (\epsilon_l^\text{pl})^{k} = (\epsilon_l^\text{pl})^{k-1} + \Delta \epsilon_l^\text{pl}, 
\quad \alpha_l^{k} = \alpha_l^{k-1} + | \Delta \epsilon_l^\text{pl} |
$$
by solving the incremental version of the evolution equation (\ref{eq:evol}) for $\{\Delta y_n\}$ and $\{ \Delta \epsilon_l^\text{pl}\}$.   Above, the last identify follows from the discretization of  (\ref{eq:plastic work}).

We first solve the evolution equation for the nodal positions by holding the linkwise plastic variables fixed.  This reduces to the equilibrium equation, 
$$
\frac{\partial {\mathcal E}}{\partial y_n} ( \{y_n^{k-1} + \Delta y_n\}, (\epsilon_l^\text{pl})^{k-1} \} ) = 0.
$$
We solve this for the $\{\Delta y_n\}$ using Newton-Raphson iteration up to a tolerance of $10^{-10}$ in the normalized Euclidean norm.  We then seek the plastic update $\{\Delta \epsilon_l^\text{pl} \}$ from the second of the evolution equations,
\begin{align} \label{eq:plastic update}
	\begin{split}
		0 & \in - K \left(\frac{|  \sum_j B_{lj} y_j^{k+1} |-L_l^0}{L_l^0} - ((\epsilon_l^\text{pl})^k + \Delta \epsilon_l^\text{pl}) \right) \\
		& \quad \quad + f_0 \left(1+ \left(\frac{\alpha_l^k+ |\Delta \epsilon_l^\text{pl}|}{\epsilon_0^p}\right)^N + \left( \frac{|\Delta \epsilon^\text{pl}_l|}{\dot \epsilon_0^\text{pl} \Delta t} \right)^m \right) 
		H(|\Delta \epsilon_l^\text{pl}|) \ \text{sgn}(\Delta \epsilon_l^\text{pl}).
	\end{split}
\end{align}
Above, $H$ is the Heaviside function that takes any value in $[0,1]$ at the origin.

%
%

\newpage

\bibliography{reference.bib}

\end{document}